\documentclass[aps,prl,reprint,twocolumn,showpacs,epsf]{revtex4}
\usepackage{graphicx}
\usepackage{psfrag}
\usepackage{color}

\def\a{\alpha}
\def\b{\beta}

\begin{document}
\bibliographystyle{apsrev}


\title{The road to catastrophe: 
stability and collapse in 2D driven particle systems}



\author{M. R. D'Orsogna, Y. L. Chuang, A. L. Bertozzi, L. S. Chayes}

\affiliation{Department of Mathematics, UCLA, Los Angeles, CA 90095}
\affiliation{Department of Physics, Duke University, Durham, NC 27708}


\date{\today}

\begin{abstract}

Understanding collective properties of 
driven particle systems is
significant for naturally occurring
aggregates and because the knowledge gained can be used as 
building blocks for the design of artificial ones. 
We model self propelling biological or artificial individuals 
interacting through pairwise attractive and repulsive forces.
For the first time, we are able to predict stability and morphology
of organization starting from the shape of the two-body 
interaction. We present a coherent theory, based on fundamental 
statistical mechanics, for all possible phases of collective motion.

\end{abstract}


\maketitle

The swarming of multi-agent systems \cite{camazine}
is a fascinating natural phenomenon.  
The patterns formed by many self-assembling species pose a 
wealth of evolutionary \cite{parrish} 
and biological \cite {couzin05, couzin02} 
questions, as 
well as structural and physical 
\cite {levine, toner, mogilner} 
ones.  In more 
recent years, understanding the operating principles of natural swarms has 
also turned into a useful tool for the intelligent 
design and control of man-made vehicles \cite{bona1, passino}.  

One of the main unresolved issues 
arising both in artificially controlled and biological swarms is the 
ability to predict stability with respect to size.  If well 
defined spacings amongst individuals exist, swarm size 
typically increases 
with particle number, in a
`crystal' like fashion.  This is generally
true in animal flocks and might be a 
desirable feature in robotic systems.
On the other hand, natural examples 
exist of swarms that shrink in size
as particle number increases.  For instance, 
in the early development of the
{\it Myxococcus xanthus} or {\it Stigmatella auriantaca}
fruiting bodies 
\cite{koch} two-dimensional bacterial vortices arise 
and grow until the vortices collapse inward, 
individual cells occupy the central 
core and a localized three dimensional structure appears.  
Although many swarming systems have been 
studied, and in some cases specific phase 
transitions have been observed 
\cite{couzin02,mogilner, passino}, 
a systematic prediction of 
whether a swarm will collapse on itself or not
as the number of constituents increases, has been lacking.

In this Letter, we apply fundamental principles from
statistical  mechanics to accurately
predict the geometry and stability of swarming systems.  
Specifically, we consider 
$N$ self-propelled particles 
powered by biological or mechanical motors,
that experience a frictional force,
leading to a preferred 
characteristic speed \cite{cdp}.  
The particles also interact
by means of a two-body generalized Morse potential.  Previous related 
work \cite{levine}, 
showed that in some cases localized vortices may form.  Here, 
we explore the {\it entire} 
phase space defined by the interaction potential and 
predict pattern geometry and stability.
The identical $N$ particles obey the equations of motion: 
\begin{eqnarray}
\label{eqom1}
\frac {\partial \vec x_i} {\partial t} &=& \vec v_i, \\
\nonumber
\end{eqnarray}
\vspace{-1.4cm}
\begin{eqnarray}
\label{eqom2}
m \frac {\partial \vec v_i} {\partial t} &=& 
(\alpha - \beta | \vec v_i|^2) \vec v_i - \vec \nabla_i U(\vec x_i),
\end{eqnarray}
\noindent
with the generalized Morse potential given as: 
\begin{eqnarray}
\label{Morse}
U(\vec x_i) &=& \sum_{j \neq i} 
[C_r e^{-|\vec x_i - \vec x_j| / \ell_r} -
C_a e^{-| \vec x_i - \vec x_j| / \ell_a}].
\end{eqnarray}
\noindent		
Here, $1\,\leq i\, \leq N$, and $\ell_a, \ell_r$ represent the range of 
the attractive and repulsive part of 
the potential and $C_a,C_r$ are their respective amplitudes.  
We remark that our 
analysis need not be confined to Morse-type potentials -- these are simply a 
mathematical convenience. Combinations of other attractive and 
repulsive potentials lead to collective behaviors similar to those seen here.  
In particular, a similar analysis can be extended to the 
interactions of Ref.\,\cite{mogilner}
and  of Ref.\,\cite{passino},  
providing a much more complete prediction of stability.

In the special case of velocity independent forces, i.e. 
$\a = \b =0$, Eqns.\,\ref{eqom1}-\ref{Morse}
form a typical Hamiltonian system with a conserved
energy. For a large number of particles, 
according to the ideas of statistical mechanics,
the behavior of such a system should be described by a finite
temperature (Maxwell-Boltzmann) distribution with the 
energy in the model determining the temperature \cite{huang}.
A Maxwell-Boltzmann description is even more applicable when
there is some mechanism for exchange of energy with the 
environment.
Here, we argue that the parameters $\a$ and $\b$ in the model 
of Eqns.\,\ref{eqom1}-\ref{Morse} provide a local 
mechanism for such energy exchanges. Thus, the gross features of our 
system should be described by a classical statistical mechanics model 
with the temperature parameter determined by $\a$ and $\b$.
Of course the detailed local behaviors, as well as the large scale 
collective dynamics, may be quite different -- and ostensibly 
more interesting -- than the corresponding system obeying
Maxwell-Boltzmann statistics.

In large systems that obey the laws of statistical mechanics it is 
expected that thermodynamics will emerge as size and number of 
constituents tend to infinity.  In order to ensure a 
smooth passage to the thermodynamic limit, the 
microscopic interactions must 
respect certain constraints.  The most important of these is 
{\it H-stability}: for a set of $N$ interacting particles,
the potential energy $U$ is said to be H-stable if a constant $B \geq 0$
exists such that $U \geq -N B $ \cite{ruelle}.
This property ensures thermodynamic behavior, in particular
that the $N$ particles will not collapse as $N \rightarrow \infty$.
Non H-stable systems are also called catastrophic.
For Morse type interactions, conditions for 
H-stability are known.  
For example, if the $d$-dimensional integral of 
the potential is negative, the system is non H-stable.
In this case,
as $N$ increases, the particles collapse
into a dense body with energy per 
particle proportional to $N$. On the other
hand, for thermodynamic systems, the 
energy per particle will be 
asymptotically constant.  
The stability phase diagram of the Morse potential is shown in Fig.\,1. 
As we shall see the H-stability or lack thereof is instrumental
in predicting the behavior of swarming systems obeying the likes
of Eqns.\,\ref{eqom1}-\ref{Morse}.

\begin{figure} [t]
\begin{center}
\includegraphics[height=1.8in]{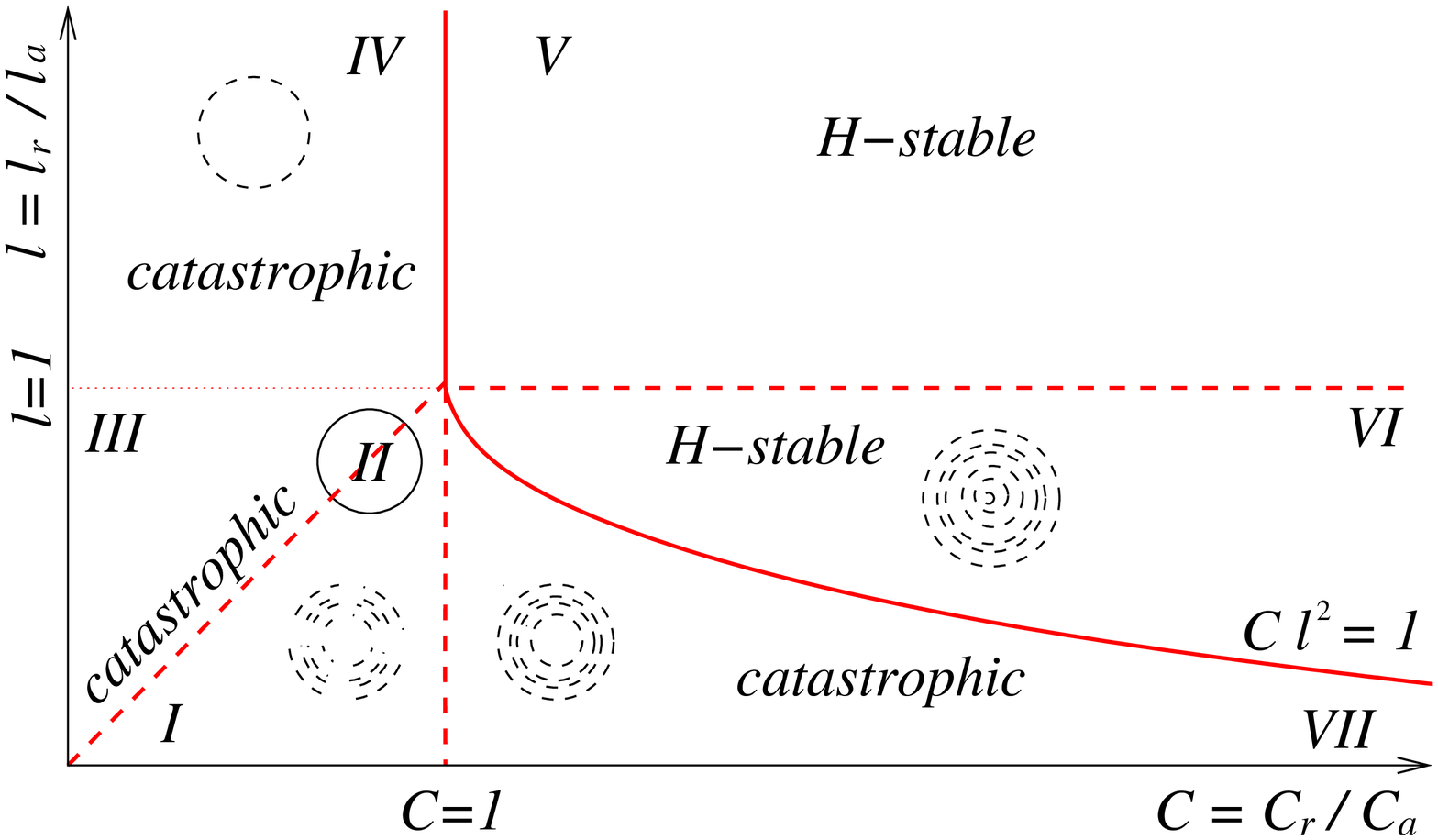}
\end{center}
\caption{H-Stability phase diagram of the Morse potential.
The catastrophic 
regions correspond to parameter ratios 
$\ell = \ell_r / \ell_a$ and $C = C_r / C_a$ for 
which the thermodynamic limit does not exist.  
Extrema of the potential 
$d_{min}$ exist only for  $\ell > \max \{1, C\}$ and for 
$\ell < \min \{1, C \}$. In these 
cases $ d_{min} =  \ell_r \, \log \,( \ell / C) / \,(\ell - 1)$.}
\label{phasediag}
\end{figure} 

For the full model, 
we numerically integrate Eqns.\,\ref{eqom1}-\ref{Morse} 
using a fourth order Adams-Bashforth 
method \cite{gene} 
for free boundaries, effectively allowing an infinite range of 
motion for the particles.  Initial conditions are chosen with localized 
particles and random velocities.  
The resulting behavior is consistent with the 
stability or catastrophic predictions of Fig.\,1.  
We discuss system behavior in each region of the phase diagram of 
Fig.\,1.

\begin{figure}[t]
\label{clump}
\includegraphics[height=5.0 in]{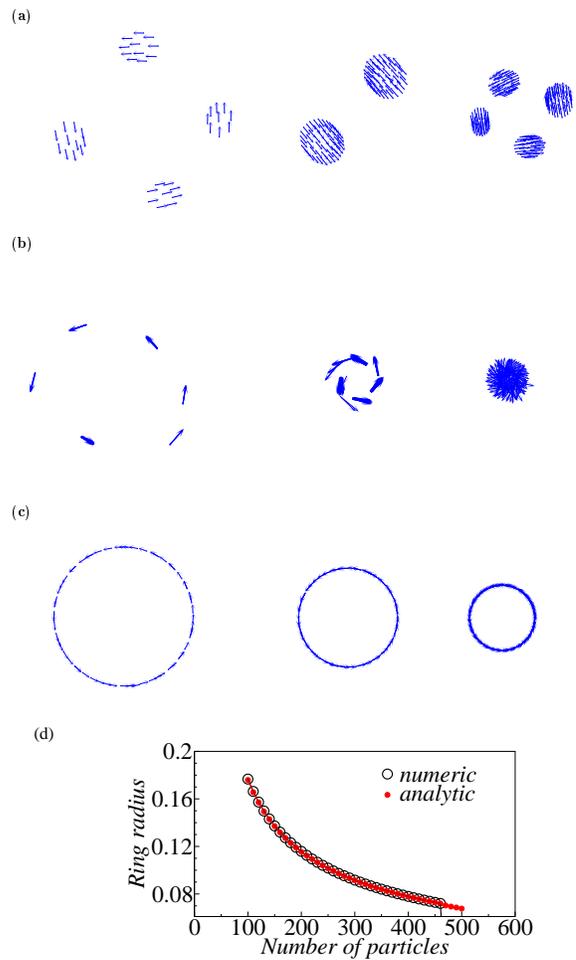}
\caption{
Catastrophic geometry.  
(a) {\it Clumps}.  
From left to right $N = 40, 100, 150$.  
Clumps coalesce as $N$ increases.
The set parameters are 
$\a = 1, \b = 0.5, C_a = \ell_a = 1, C_r = 0.6, \ell_r = 0.5$.  
(b)  {\it Ring clumping}.  
From left to right $N = 40, 100, 200$.  
Parameters are the same as in Fig.\,2a, but 
$\ell_r = 1.2$. 
(c) {\it Rings}.  From left to right 
$N = 60, 100, 200$.  
Parameters are the same as in Fig.\,2a, but 
$C_r = 0.5$.  
(d)  
Ring radius as a function of $N$ from numerical data 
and Eq.\,\ref{ring}.  
Parameters are the same as in Fig.\,2c. 
Fitting data from Eq.\,\ref{ring} yields $R \sim N^{-0.52}$. }
\end{figure}

In the case $\min \{C, \ell \} < 1$ the interparticle potential 
is of catastrophic nature 
and globally attractive.  For $\b \neq 0$, particles tend to sustain 
a constant speed $|\vec v_i|^2 \sim \a/\b$, 
while subject 
to attractive forces.  This competition leads to 
non-equilibrium configurational patterns.  
We distinguish three subregions in the 
$\min \{C, \ell \} < 1$ region:  
$\{\ell  < C\}, \{\ell  = C\}$ and $\{\ell  > C\}$, respectively 
regions I, II and III of Fig.\,1.  
In region I a potential minimum $d_{min}$ exists and the $N$ 
particles self-organize by creating 
multi-particle clumps.  
Within each clump 
the particles travel parallel to each other defining a collective direction.  
Because of the rotational velocity, this direction 
changes in time and the clumps rotate about their center
of mass (Fig.\,2a).  
Catastrophic behavior is evident in the fact that as $N$ 
increases, the clumped structures shrink instead of swelling.  
Interparticle distance also becomes smaller and eventually, 
as  $N \rightarrow \infty$ clumps lose their coherence
and merge.  
The bisectant $\{\ell  = C\}$, region II, is the borderline for the 
existence of extrema.   
Here, the potential minimum occurs for $d_{min}  = 0$, no 
associated finite length scale exists, and rings are developed 
(Fig.\,2c).  
Assuming equidistant particle spacing, the ring radius $R$ may be 
estimated by balancing the centrifugal and centripetal forces.  
An approximate implicit expression for $R$ is given by:
\begin{eqnarray}
\label{ring}
\frac {\a} {2 R \b} = \sum_{n=1}^{N/2} 
\sin \left(\frac {\pi n} N \right)
\,\left[ \frac 
{C_a} {\ell_a} e^{- \frac {2 R}{\ell_a} \sin 
\left[ \frac{\pi n } {N} \right]} -
\frac {C_r} {\ell_r} e^{- \frac {2 R} {\ell_r} \sin \left[\frac{\pi n} 
{N} \right]} \right] \\
\nonumber
\end{eqnarray}
\noindent
Estimates of $R$ as given by Eq.\,\ref{ring} 
match extremely well those obtained numerically
as seen in Fig.\,2d. Circular structures are also seen
in the swarms of Ref.\,\cite{ebeling}.
For $\{\ell  > C\}$ in region 
III of Fig.\,1, clumped structures appear although 
there is no minimum in the potential.  
In particular, no intrinsic 
interparticle spacing exists and the clumps consist of 
superimposed particles traveling along a ring: 
this type of collective motion is energetically 
more favorable than uniform spacing among particles.  
An example of ring clumping is shown in Fig.\,2b.

A clumped ring structure also 
appears in the $\{C < 1 < \ell\}$ regime of region IV.  
The observed behavior 
is very similar to what described in the $\{\ell  > C \}$ 
case above,  with the 
difference that here the potential defines a maximum, and for low 
particle numbers the extra constraint of avoiding energetically 
costly interparticle spacings has to be considered.  Region V 
of Fig.\,1 where $\max \{\ell ,C\} > 1$ corresponds to the H-stable regime. 
Here, the interparticle potential is characterized by overall 
repulsive behavior and is minimized by infinite separation.  Thus, 
as $N \rightarrow \infty$ the particles 
will tend to occupy the entire volume at their 
disposal.  The entire region is a `gaseous' 
phase with particle speed peaked at 
$|\vec v_i|^2 = \a/\b$.  

The most interesting region of the phase diagram is defined by 
$\{\ell < 1 < C\}$, regions VI and VII of the phase diagram.  
Here, the potential is characterized by short range repulsion and long 
range attraction.  A potential minimum exists and defines a length scale 
$d_{min}$.  
The $C \ell^2 =1$ curve of Fig.\,1 parts the 
thermodynamically stable region 
VI from the thermodynamically catastrophic region VII.  
Although the main features of the two-body potentials are similar, 
different H-stability properties lead 
to very different self-organizational 
behaviors in the moderate and large particle limits. 
 
\begin{figure}
\includegraphics[height=1.2 in]{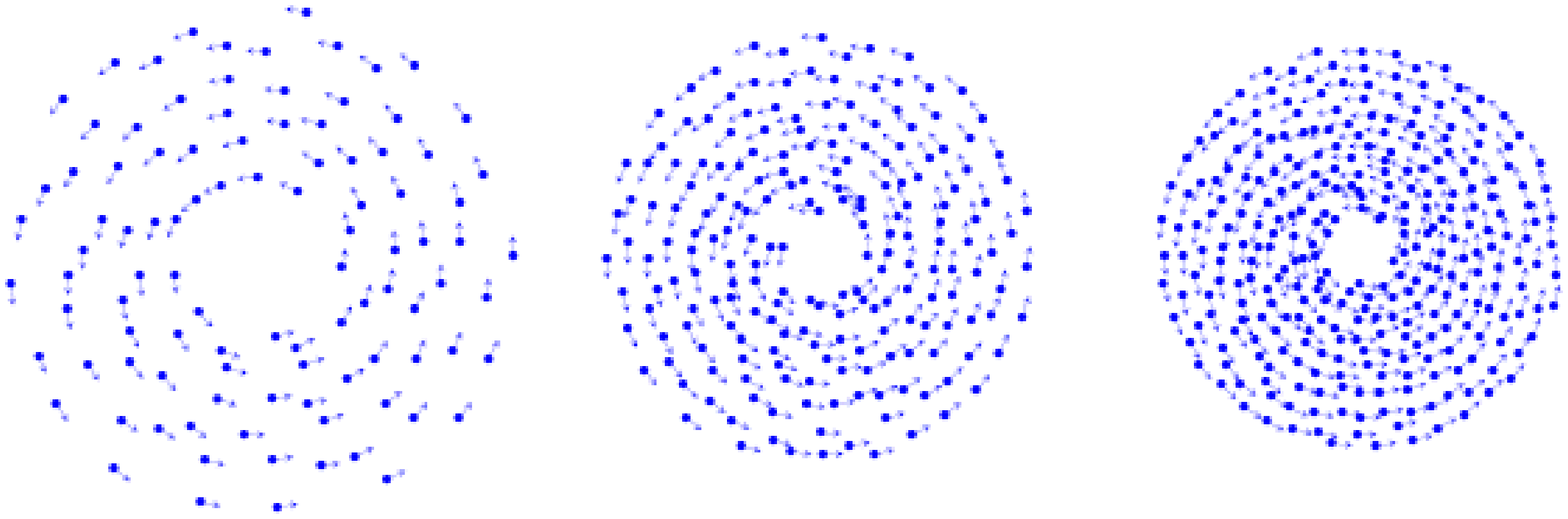}
\caption{
Snapshots of swarms for different values of $N$ in the 
catastrophic regime defined by region VII of Fig.\,1.  From left 
to right $N = 100, 200, 300$. Note 
the decrease in the vortex area and the dramatic density increase.  
The chosen parameters are: 
$C_a = 0.5, C_r = 1, \ell_a = 2,  \ell_r = 0.5.$  
Self-propulsion and friction are fixed at 
$\a =1.6$ and $\b = 0.5$.}
\end{figure}

\begin{figure} [t]
\label{area}
\includegraphics[height=2.6in]{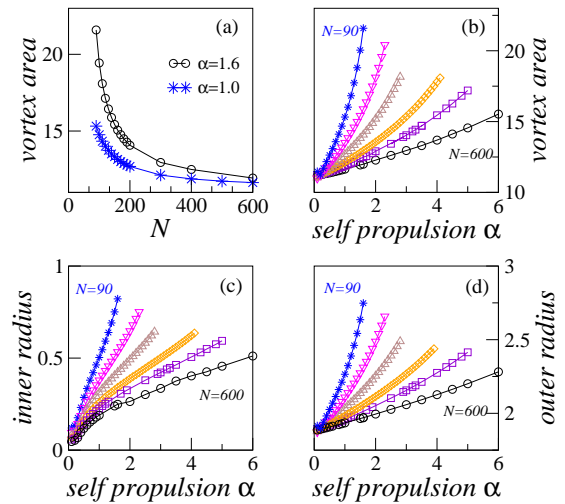}
\caption{
Vortex scalings for the catastrophic Morse potential.  
The parameters are set as in Fig.\,3.  The friction term 
$\b = 0.5$.  
(a) 
Vortex area as a function of $N$ for $\a=1.0, 1.6$.  
Note the dramatic decrease with $N$.  (b)
Vortex area as a function of (a) for various $N$.  
From top to bottom $N = 90, 140, 200, 300, 400, 600$.  
For any fixed $\a$ the vortex area decreases with 
$N$.  (c) Inner and (d)
Outer radii of the catastrophic vortices as a function of $\a$.  
The particle numbers are the same 
as in Fig.\,4b.  Both radii increase with $\a$ but decrease with 
$N$.  For large $N$ the inner core disappears. }
\end{figure}

Region VI with $\{1/ \sqrt{C} < \ell < 1\}$ 
corresponds to thermodynamic 
stability.  At finite $N$, 
and for various values of $\a/\b$, 
particles approach the characteristic velocity 
$|v_i^2|= \a/\b$ and reach 
a kinetic energy much greater than the confining interaction potential.  
The swarming agents tend to disperse as individuals.  
For much smaller of values of 
$\a/\b$ , the $N$ particles assemble into organized 
structures with well defined spacings, which in the 
large particle limit tend to a finite value.  
Particles will then either 
swarm coherently in a rigid disk aggregate or flock with a finite 
center of mass velocity, depending on the initial conditions.  In both cases, 
the motion is rigid body-like and interparticle distances are preserved.  
For $\a/\b \rightarrow 0$, the particles assemble into static, 
locally crystalline structures.  

\begin{figure} [t]
\begin{center}
\includegraphics[height=1.8in]{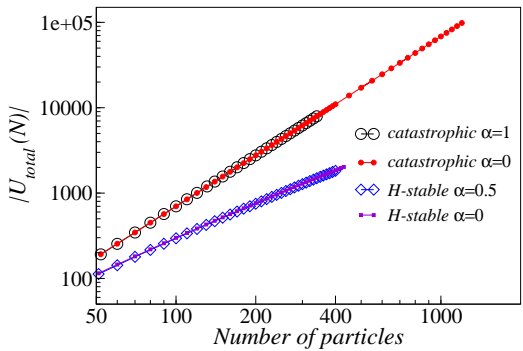}
\end{center}
\caption{
Absolute value of the total Morse potential energy,
$|U_{total}|$, as a function of $N$. 
Several choices of $\a$ are shown with fixed 
$\b = 0.5$. 
The upper curves correspond to the catastrophic 
regime with the same Morse parameters as in Fig.\,3. 
The scaling law is $ - U_{total} \sim N^{2.0}$ 
for both $\a=1$ and $\a=0$.  
The lower curves correspond to the H-stable parameters 
$C_a = 0.5, C_r = 1$ and $\ell_a  =  2, \ell_r  = 1.5$. 
The scaling law is $-U_{total} \sim N^{1.0}$ for both
$\a=0$ and $\a=0.5$.
Scaling is linear in the H-stable case,
and quadratic in the catastrophic one.}
\label{totalenergy}
\end{figure}

Region VII where 
$\{\ell < 1/ \sqrt{C} < 1\}$, corresponds to 
thermodynamic instability; all cases
examined in Ref.\,\cite{levine}
concern this region. As in the previous case, 
for finite $N$, large values 
of $\a/\b$ will lead to a gaseous phase and 
very small values to crystalline structures 
whose motion is rigid body-like.  However, quite unlike 
the H-stable scenario discussed above, these structures are unstable 
with respect to particle number, 
and in the $N \rightarrow \infty$ 
limit will collapse. 
At intermediate values of 
$\a/\b$, vortex structures appear 
with particles traveling close to the characteristic
speed $| \vec v_i| ^2 \sim \a/ \b$.  
Here, vortex size  {\it decreases} 
dramatically as a function of particle number as seen in Figs.\,3,\,4.  
Also, for finite $N$, vortices rotating
counter-clockwise and clockwise may coexist, depending 
on the initial conditions.  
In this regime, the occurrence of double spiraling is 
visually most dramatic since it occurs within vortices, however double 
spiraling is a feature of the entire catastrophic part of the phase 
diagram and coexisting left and right direction of motions for clumped or 
equispaced rings occur as well.  Double spirals are thus 
a strong indication 
of the non H-stable nature of the potential.
Another typical feature of the catastrophic regime is
that energy per particle does not asymptotically reach a constant value.
This is seen in Fig.\,\ref{totalenergy} 
where, in the non H-stable regime,
the total energy scales quadratically,
so that energy per particle grows linearly.
Here, interparticle separation (not shown) decreases dramatically
as $N \rightarrow \infty$.
For comparison, in the H-stable regime, 
the total potential energy scales linearly with $N$ 
and energy per particle 
does approach a constant. Likewise, as $N \rightarrow \infty$ 
interparticle separation 
is asymptotically constant.

The observed swarming behavior of {\it M. xanthus} 
is consistent with the  
catastrophic regime VII of the Morse potential where core free vortex 
structures can arise.  
The absence of a hard component for the interparticle interaction is 
justified by the fact that {\it M. xanthus} 
cells can penetrate each other by 
crawling.  We propose the following qualitative and coarse scenario for 
the initial stages of aggregation.  As the number of constituents increases, 
so does particle speed $\sqrt {\a/\b}$.   
This is consistent with the observed enhancement of C-signal 
activity among particles \cite{jelsbak} that increases motility.  
The bacterial vortex then increases its size with $N$ and double spirals 
may coexist, as reported in the literature \cite{oster1}.  
Eventually, bacterial speed reaches an upper limit and increasing $N$ 
will lead the vortex to collapsed behavior, until,
finally, it evolves into a complex 
3D structure \cite{kuner} due to finite height of the cells and other 
features not accounted for in Eqns.\ref{eqom1}-\ref{Morse}.  

Simple modeling of interacting particles combined with 
thermodynamic reasoning can account for complex and subtle phenomena in 
biological systems. Furthermore, in spite of its name, the existence of 
a catastrophic regime might be of great benefit to the development of 
unmanned vehicle technology for the added versatility it offers.  
`Soft core' robot interactions could be implemented by specialized 
cooperations at short distances e.g. robots crawling over or rotating 
about each other or by setting parameters so that actual 
individual size is miniscule compared to relevant length scales. 
For large or even moderate number of vehicles the programming of a 
crossover from an H-stable to a catastrophic 
regime could lead the robots to change 
from a dispersive 
(searching) behavior to convergence at a specified site. 

We thank H. Levine and D. Marthaler for useful discussions. 
We acknowledge support from ARO and NSF through grants W911NF-05-1-0112 
and DMS-0306167.

 \newpage



\end{document}